\documentclass{article}

\usepackage{neurips_2023}

\usepackage[utf8]{inputenc} 
\usepackage[T1]{fontenc}    
\usepackage{hyperref}       
\usepackage{url}            
\usepackage{booktabs}       
\usepackage{amsfonts}       
\usepackage{nicefrac}       
\usepackage{microtype}      
\usepackage{xcolor}         
\usepackage{algorithm}
\usepackage{setspace}
\usepackage{algpseudocode}
\usepackage{amsmath}
\usepackage{graphicx} 
\usepackage{tabularx}

\usepackage{multirow}
\usepackage{tikz}

\newcommand\encircle[1]{%
    \tikz[baseline=(X.base)] 
    \node (X) [draw, shape=circle, inner sep=0, fill=black, text=white, scale=0.8] {\strut #1};%
}
\usepackage{setspace}
\usepackage{amsmath}
\usepackage{sectsty}

\title{Scalable Readability Evaluation for Graph Layouts: 2D Geometric Distributed Algorithms}

\author{%
  Sanggeon Yun \\
  University of California, Irvine\\
  sanggeoy@uci.edu\\
}

\begin{document}

\maketitle


\section{Introduction}

Graphs, consisting of vertices and edges, are vital for representing complex relationships in fields like social networks, finance, and blockchain~\cite{henry2007matlink, li2015visualizing, lin2015integer, chang2007wirevis, niu2018visual, maccas2020vabank, mcginn2016visualizing}. Visualizing these graphs helps analysts identify structural patterns, with readability metrics—such as node occlusion and edge crossing—assessing layout clarity~\cite{ke2004major}. However, calculating these metrics is computationally intensive, making scalability a challenge for large graphs~\cite{klammler2018aesthetic,gove2018pays}. Without efficient readability metrics, layout generation processes—despite numerous studies focused on accelerating them~\cite{godiyal2008rapid,frishman2007multi,mi2016interactive,brinkmann2017exploiting, hinge2015distributed,arleo2017large,hinge2017mugdad,gomez2018visualizing}—face bottleneck, making it challenging to select or produce optimized layouts swiftly. Previous approaches attempted to accelerate this process through machine learning models. Machine learning approaches~\cite{haleem2019evaluating} aimed to predict readability scores from rendered images of graphs. While these models offered some improvement, they struggled with scalability and accuracy, especially for graphs with thousands of nodes. For instance, this approach requires substantial memory to process large images, as it relies on rendered images of the graph; graphs with more than 600 nodes cannot be inputted into the model, and errors can exceed 55\% in some readability metrics due to difficulties in generalizing across diverse graph layouts.
This study addresses these limitations by introducing scalable algorithms for readability evaluation in distributed environments, utilizing Spark’s DataFrame~\cite{armbrust2015spark} and GraphFrame~\cite{dave2016graphframes} frameworks to efficiently manage large data volumes across multiple machines. Experimental results show that these distributed algorithms significantly reduce computation time, achieving up to a 17$\times$ speedup for node occlusion and a 146$\times$ improvement for edge crossing on large datasets. These enhancements make scalable graph readability evaluation practical and efficient, overcoming the limitations of previous machine-learning approaches.

\section{Background}

\subsection{Readability Metrics}\label{s:sec2.1}
Several readability metrics~\cite{purchase2002metrics,dunne2015readability} help evaluate the clarity of graph layouts, allowing for quantitative comparisons of their aesthetic quality. This study focuses on optimizing five key readability metrics in distributed environments.

\begin{itemize}
    \item \textbf{Node Occlusion}: This measures overlapping nodes. Two nodes are considered occluded if the distance between them is less than a defined diameter, requiring an $O(|V|^2)$ complexity where $V$ is the set of vertices.

    \item \textbf{Minimum Angle}: This metric calculates how close the angles between connected edges are to an ideal minimum. It involves sorting and computing angle differences, with a complexity of $O\left(\sum_{v \in V} |c(v)| \log{|c(v)|}\right)$ where $c(v)$ represents edges connected to vertex $v$.

    \item \textbf{Edge Length Variation}: This measures how much edge lengths deviate from their average, indicating uniformity. It has a complexity of $O(|E|)$, where $E$ is the set of edges.

    \item \textbf{Edge Crossing}: This metric counts intersecting edge pairs, with fewer crossings indicating less clutter. The complexity is $O(|E|^2)$.

    \item \textbf{Edge Crossing Angle}: This calculates the average difference between the actual crossing angles of edges and an ideal angle, usually 70 degrees~\cite{huang2008effects}, with a complexity also of $O(|E|^2)$.
\end{itemize}

\subsection{Spark’s DataFrame and GraphFrames Framework}

Spark~\cite{zaharia2016apache} is an open-source platform for large-scale data processing, known for being faster than MapReduce~\cite{dean2008mapreduce}. Its core data structure, the Resilient Distributed Dataset (RDD), enables parallel computation. DataFrames in Spark are an abstraction of RDDs, representing data in a table-like format. Spark’s DataFrame~\cite{armbrust2015spark} API offers operations like:

\begin{itemize}
    \item \textbf{Join}: Combines two DataFrames based on shared columns, requiring partition alignment, which can be computationally expensive.
    \item \textbf{Explode}: Separates array elements into individual rows.
    \item \textbf{GroupBy}: Groups rows by specified columns, enabling aggregate operations.
    \item \textbf{Aggregate}: Supports built-in and user-defined functions for aggregating data, often used after \texttt{GroupBy}.
    \item \textbf{Distinct}: Removes duplicate rows.
    \item \textbf{Count}: Returns the number of rows in a DataFrame.
\end{itemize}

GraphFrames, an extension of Spark, supports graph-parallel computations, offering functions such as \texttt{aggregateMessages}, which aggregates messages for each vertex.

\subsection{Distributed Graph Layout Algorithms}\label{s:sec2.3}

Several distributed algorithms focus on graph layout generation~\cite{gomez2018visualizing, arleo2017large}. The Fruchterman-Reingold algorithm\cite{fruchterman1991graph} uses attractive and repulsive forces between nodes to determine positions, while GiLA~\cite{arleo2017large} and Multi-GiLA~\cite{arleo2018distributed} use Giraph to process large graphs by approximating these forces. GiLA calculates forces between each vertex and its neighbors, while Multi-GiLA expands on this to handle large-scale graphs cost-effectively on distributed cloud platforms.

\section{Distributed Readability Evaluation Algorithm}
    \label{sec:algo}
    \subsection{Exact Algorithm} \label{s:sec3.1}
        We introduce the five readability metrics that we implemented in Spark to be run on distributed environment. Exact algorithms are designed to compute readability metrics in a straightforward approach without any approximation by fully utilizing DataFrame and GraphFrames APIs.

        \subsubsection{Distributed Node Occlusion} \label{s:sec3.1.1}
            The simplest approach to compute $N_c$ is to compare all the vertices. We used Spark dataframe's \textit{join} operation to achieve this. Specifically, the algorithm generates two dataframes $D_{pos1}$ and $D_{pos2}$ which are identical to the $D_{pos}$, but with different column names. Here, the dataframe $D_{pos}$ contains the ids and $xy$-coordinates of vertices, and the radius of the boundary ($r$). Next, it performs the \textit{join} operation with two conditions: 1) the order of vertex ids, and 2) euclidean distance. With the first condition, it prevents having duplicates where two rows with the same vertices paired in a different order. The second condition ensures that each vertex joins with the vertices whose boundaries are overlapping. The steps for getting node occlusion are presented in \autoref{algorithm:1}.

\begin{spacing}{1.5}
\begin{algorithm}
\caption{Distributed node occlusion}\label{algorithm:1}
\hspace*{\algorithmicindent} \textbf{Input:}\\
\hspace*{\algorithmicindent}\hspace*{\algorithmicindent}$D_{pos}$: A dataframe containing vertex ids and $(x, y)$ positions\\
\hspace*{\algorithmicindent}\hspace*{\algorithmicindent}$r$: Radius of boundary circle\\
\hspace*{\algorithmicindent} \textbf{Output:} Node occlusion
\begin{algorithmic}[1]
\Procedure{DistributedNodeOcclusion}{$D_{pos},r$}
    \State $D_{pos1} \leftarrow $ $D_{pos}$ with column renamed $(v, pos1)$
    \State $D_{pos2} \leftarrow $ $D_{pos}$ with column renamed $(u, pos2)$
    \State $N_c \leftarrow$ number of rows of dataframe: $D_{pos1}$ join $D_{pos2}$ with $v < u$ and $||pos1 - pos2||^2 < (2r)^2$ condition \Comment{node occlusion}
\EndProcedure
\end{algorithmic}
\end{algorithm}
\end{spacing}

        \subsubsection{Distributed Minimum Angle} \label{s:sec3.1.2}
            With given dataframes $D_{pos}$ containing vertex ids and their $xy$-coordinates and $D_e$ containing edge list, the algorithm first initializes a \textit{GraphFrame} object. Then, to find the minimum angle for each vertex, it collects angles $a_i\in[0, 2\pi]$ that are formed with $x$-axis for all edges that are connected to each vertex by using the \textit{aggregateMessages} operation. As a result of the previous step, it now has a dataframe $D_a$ having array of angles for each vertex. Based on $D_a$, it creates a new column containing $\frac{\phi(\nu) - \phi_{min}(\nu)}{\phi(\nu)}$ for each vertex $\nu$. $\phi(\nu)$ is easily induced using the length of the array. $\phi_{min}(\nu)$ is computed by sorting the given array in non-decreasing order and calculating the difference between neighboring angles including the difference between the first element and the last element in the sorted array. We can notice that the minimum difference value in the array is equal to the value of $\phi_{min}(\nu)$. Finally, the value of $M_a$ is computed by applying \textit{aggregate} to the newly generated column in $D_a$. The steps for getting the minimum angle are presented in \autoref{algorithm:2}.

            \begin{spacing}{1.5}
            \begin{algorithm}
            \caption{Distributed minimum angle}\label{algorithm:2}
            \hspace*{\algorithmicindent} \textbf{Input:}\\
            \hspace*{\algorithmicindent}\hspace*{\algorithmicindent}$D_{pos}$: A dataframe containing vertex ids and $(x, y)$ positions\\
            \hspace*{\algorithmicindent}\hspace*{\algorithmicindent}$D_{e}$: A dataframe containing edge list\\
            \hspace*{\algorithmicindent} \textbf{Output:} Minimum angle
            \begin{algorithmic}[1]
            \Procedure{DistributedMinimumAngle}{$D_{pos},D_{e}$}
                \Function{GetMinAngle}{$\{a_1, a_2, ..., a_n\}$}
                    \State sort $a_i$ in non-decreasing order
                    \State $\Delta \leftarrow \{2\pi - a_n + a_1\}$
                    \For{$i\in \{2, 3, ..., n\}$}
                        \State $\Delta \leftarrow \Delta \cup \{a_i - a_{i-1}\}$
                    \EndFor
                    \State \Return{$\min{\Delta}$}
                \EndFunction
                \State $G_{a} \leftarrow \text{GraphFrame}(D_{pos}, D_{e})$
                \State $D_{a} \leftarrow G_{a}.\text{aggregateMessages}$ \Comment{collect angles $a^v_i\in [0, 2\pi]$ for each vertex $v$}
                \State $D_{a} \leftarrow D_{a}$ with new column $d_v = \frac{2\pi / |a^v_i| - \text{\Call{GetMinAngle}{$a^v_i$}}}{2\pi / |a^v_i|}$ 
                \State $M_a \leftarrow$ aggregate $D_{a}$ for all $d_v$ to get $1 - \sum_{v\in D_{pos}}{d_v}$ \Comment{minimum angle}
            \EndProcedure
            \end{algorithmic}
            \end{algorithm}
            \end{spacing}

        \subsubsection{Distributed Edge Length Variation} \label{s:sec3.1.3}
            Similar to the minimum angle algorithm, it also initializes a \textit{GraphFrame} object using the same dataframes. It collects the length of edges that are connected to each vertex using the \textit{aggregateMessages} operation. This generates a new dataframe $D_l$ containing an array of collected lengths of edges for each vertex. Next, it applies the \textit{explode} operation to the column containing a collection of edge lengths. Now, it computes $N_e$ and $l_\mu$ using \textit{count} operation and \textit{aggregate} operation, respectively. Finally, $\sqrt{\sum_{e\in E}{(l_e-l_\mu)^2/(N_e\times l^2_\mu)}}$ is computed using \textit{aggregate} operation with $N_e$ and $l_\mu$. By dividing it by $\sqrt{N_e-1}$, it can directly induce the value of $M_l$. The steps for getting edge length variation are presented in \autoref{algorithm:3}.
            
            \begin{spacing}{1.5}
            \begin{algorithm}
            \caption{Distributed edge length variation}\label{algorithm:3}
            \hspace*{\algorithmicindent} \textbf{Input:}\\
            \hspace*{\algorithmicindent}\hspace*{\algorithmicindent}$D_{pos}$: A dataframe containing vertex ids and $(x, y)$ positions\\
            \hspace*{\algorithmicindent}\hspace*{\algorithmicindent}$D_{e}$: A dataframe containing edge list\\
            \hspace*{\algorithmicindent} \textbf{Output:} Edge length variation
            \begin{algorithmic}[1]
            \Procedure{DistributedEdgeLengthVariation}{$D_{pos},D_{e}$}
                \State $G_{e} \leftarrow \text{GraphFrame}(D_{pos}, D_{e})$
                \State $D_{l} \leftarrow G_{e}.\text{aggregateMessages}$ \Comment{for each vertex $v$, collect length $l^v_e$ of edge $e$ connected to $v$}
                \State $D_l \leftarrow$ $D_l$ with explode operation on column $l^v_e$
                \State $N_e \leftarrow$ number of rows of dataframe $D_l$ \Comment{$N_e = |E|$}
                \State $l_\mu \leftarrow$ aggregate $D_{l}$ for all $l_e$ to get $\frac{1}{N_e}\sum_{e\in E}l_e$
                \State $l_a \leftarrow$ aggregate $D_{l}$ for all $l_e$ to get $\sqrt{\sum_{e\in E}{(l_e-l_\mu)^2/(N_e\times l^2_\mu)}}$
                \State $M_l \leftarrow \frac{l_a}{\sqrt{N_e - 1}}$ \Comment{edge length variation}
            \EndProcedure
            \end{algorithmic}
            \end{algorithm}
            \end{spacing}

        \subsubsection{Distributed Edge Crossing} \label{s:sec3.1.4}
            \begin{spacing}{1.5}
            \begin{algorithm}
            \caption{Distributed edge crossing}\label{algorithm:4}
            \hspace*{\algorithmicindent} \textbf{Input:}\\
            \hspace*{\algorithmicindent}\hspace*{\algorithmicindent}$D_{pos}$: A dataframe containing vertex ids and $(x, y)$ positions\\
            \hspace*{\algorithmicindent}\hspace*{\algorithmicindent}$D_{e}$: A dataframe containing edge list\\
            \hspace*{\algorithmicindent} \textbf{Output:} Edge crossing
            \begin{algorithmic}[1]
            \Procedure{DistributedEdgeCrossing}{$D_{pos},D_e$}
                \Function{CCW}{$\vec{A}, \vec{B}, \vec{C}$}
                    \State $c \leftarrow \overrightarrow{CA}\times \overrightarrow{AB}$ \Comment{outer product}
                    \If{$c > 0$}
                        \State \Return{$1$}
                    \ElsIf{$c < 0$}
                        \State \Return{$-1$}
                    \EndIf
                    \State \Return{$0$}
                \EndFunction
                \State $D_{epos} \leftarrow$ $D_e$ with $(x, y)$ posisions of each vertex by joining with $D_{pos}$
                \State $D_{e1} \leftarrow $ $D_{epos}$ with column renamed $(v_1, v_{pos1}, u_1, u_{pos1})$
                \State $D_{e2} \leftarrow $ $D_{epos}$ with column renamed $(v_2, v_{pos2}, u_2, u_{pos2})$
                \State $E_c \leftarrow$ number of rows of dataframe: $D_{e1}$ join $D_{e2}$ with $(v1, u1) < (v2, u2)$ and $\Call{CCW}{v_{pos1}, u_{pos1}, v_{pos2}}\times \Call{CCW}{v_{pos1}, u_{pos1}, u_{pos2}} \leq 0$ and $\Call{CCW}{v_{pos2}, u_{pos2}, v_{pos1}}\times \Call{CCW}{v_{pos2}, u_{pos2}, u_{pos1}} \leq 0$ condition \Comment{edge crossing}
            \EndProcedure
            \end{algorithmic}
            \end{algorithm}
            \end{spacing}
            To compute $E_c$, we can inspect whether a pair of edges crosses each other. 
            This can be computed by the \textit{join} operation of the Spark dataframe. With given dataframes $D_{pos}$ and $D_{e}$, the algorithm generates a new dataframe $D_{epos}$ by joining $D_e$ with $D_{pos}$ to position each vertex's $xy$-coordinate in the same row. Similar to the distributed node occlusion, it generates two dataframes $D_{epos1}$ and $D_{epos2}$ which are identical to the $D_{epos}$ but having different column names to perform \textit{join} operation. The \textit{join} operation between $D_{epos1}$ and $D_{epos2}$ is conducted with two conditions: 1) order of edge ids and 2) intersecting condition. The first condition prevents duplicate cases. It can be also implemented using vertex ids by comparing pairs of vertex ids instead of edge ids. The second condition ensures that each edge joins with edges that intersect each other. To determine whether two edges intersect or not, it uses the orientation-determining algorithm of three points also known as the CCW algorithm. For the ease of implementation, we did not consider the case where two edges are located collinearly. Finally, the \textit{count} operation is applied to the joined dataframe to result in $E_c$. The steps for getting edge crossing are presented in \autoref{algorithm:4}.

        \subsubsection{Distributed Edge Crossing Angle} \label{s:sec3.1.5}
            Edge crossing angle also requires computing crossing edges. Therefore, it uses the same procedure as the edge crossing algorithm to generate a new dataframe $D_c$ containing pairs of edges that are intersecting each other including corresponding $xy$-coordinates for each edge. After $D_c$ is generated, the algorithm creates a new column containing intersecting angles $a_c$. They can be induced using their $xy$-coordinates and $\arctan$ function. Then, the \textit{aggregate} operation is applied to the newly created column for computing the mean value of $\frac{\vartheta - a_c}{\vartheta}$. Using the aggregated value, the value of $E_{ca}$ is directly induced. The steps for getting edge crossing angle are presented in \autoref{algorithm:5}. Note that  the $CCW$ function is omitted since it is identical to the function in \autoref{algorithm:4}.
            
            \begin{spacing}{1.5}
            \begin{algorithm}
            \caption{Distributed edge crossing angle}\label{algorithm:5}
            \hspace*{\algorithmicindent} \textbf{Input:}\\
            \hspace*{\algorithmicindent}\hspace*{\algorithmicindent}$D_{pos}$: A dataframe containing vertex ids and $(x, y)$ positions\\
            \hspace*{\algorithmicindent}\hspace*{\algorithmicindent}$D_{e}$: A dataframe containing edge list\\
            \hspace*{\algorithmicindent}\hspace*{\algorithmicindent}$\vartheta$: Ideal angle\\
            \hspace*{\algorithmicindent} \textbf{Output:} Edge crossing angle
            \begin{algorithmic}[1]
            \Procedure{DistributedEdgeCrossingAngle}{$D_{pos},D_e,\vartheta$}
                \State $D_{epos} \leftarrow$ $D_e$ with $(x, y)$ posisions of each vertex by joining with $D_{pos}$
                \State $D_{e1} \leftarrow $ $D_{epos}$ with column renamed $(v_1, v_{pos1}, u_1, u_{pos1})$
                \State $D_{e2} \leftarrow $ $D_{epos}$ with column renamed $(v_2, v_{pos2}, u_2, u_{pos2})$
                \State $D_c \leftarrow $ $D_{e1}$ join $D_{e2}$ with $(v1, u1) < (v2, u2)$ and $\Call{CCW}{v_{pos1}, u_{pos1}, v_{pos2}}\times \Call{CCW}{v_{pos1}, u_{pos1}, u_{pos2}} \leq 0$ and $\Call{CCW}{v_{pos2}, u_{pos2}, v_{pos1}}\times \Call{CCW}{v_{pos2}, u_{pos2}, u_{pos1}} \leq 0$ condition \Comment{edge crossing}
                \State $D_c \leftarrow $ $D_c$ with new column $a_c$ containing intersecting angles
                \State $E_{ca} \leftarrow $ aggregate $D_c$ for all $a_c$ to get mean of $\frac{|\vartheta - a_c|}{\vartheta}$
                \State $E_{ca} \leftarrow 1 - E_{ca}$ \Comment{edge crossing angle}
            \EndProcedure
            \end{algorithmic}
            \end{algorithm}
            \end{spacing}

        \subsection{Enhanced Algorithm} \label{s:sec3.2}            
            \begin{figure}
                \centering
                \includegraphics[width=1.\columnwidth]{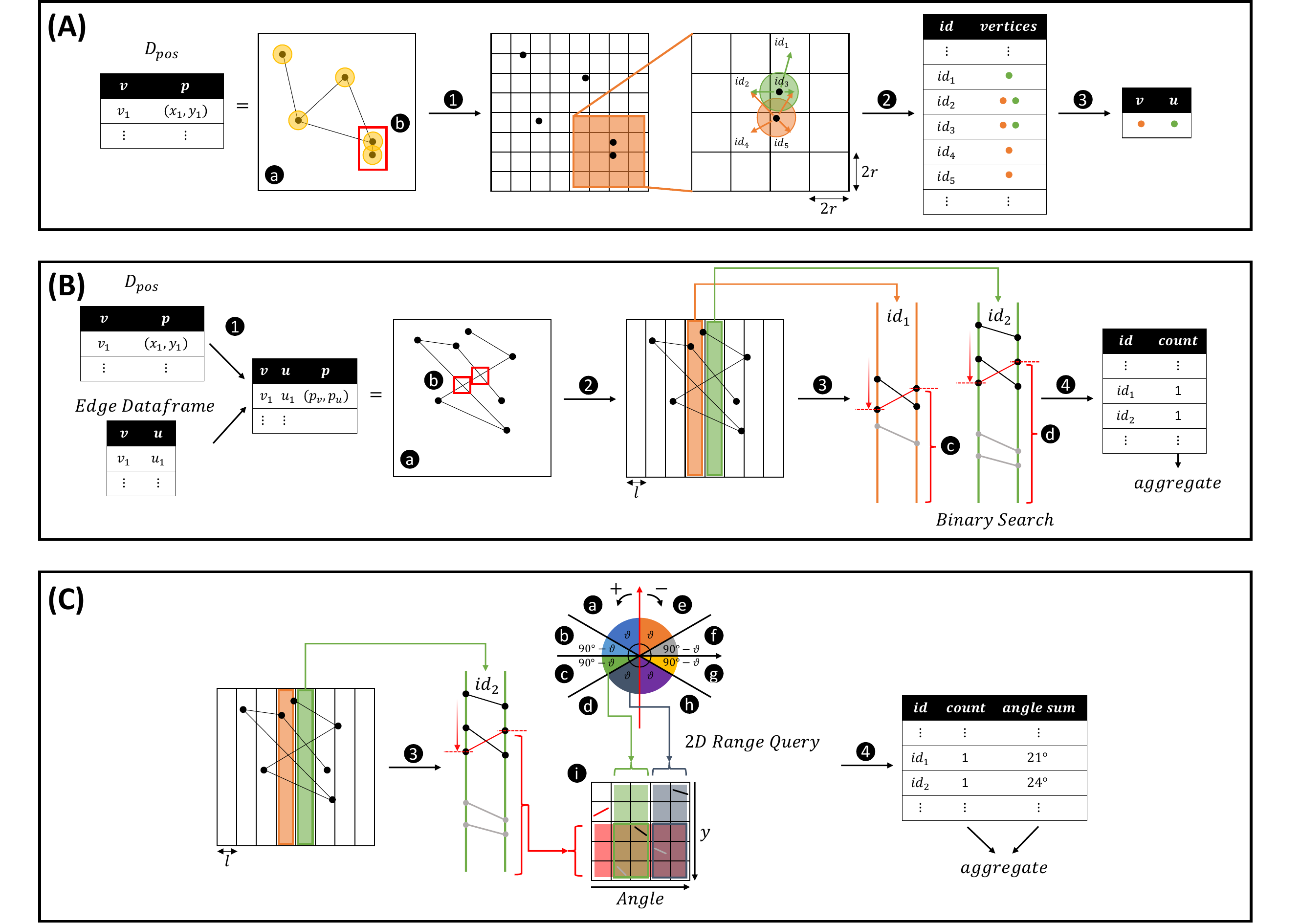}
                \caption{Overview of enhanced readability evaluation algorithms. (A) Node occlusion overview. (B) Edge crossing overview. (C) Edge crossing angle overview. Each number in the circle indicates each step of the algorithm. Note that the first two steps of the edge crossing angle are omitted since they are the same as the first two steps of edge crossing.}\label{fig:1}
            \end{figure}
            
            The most significant time-consuming task from the previous implementation is the \textit{join} operation. The \textit{join} operation with a large number of rows requires an expensive shuffle operation which includes partition transferring with each machine. This is not efficiently computed even with a large number of machines due to network latency. To avoid this, we propose enhanced readability evaluation algorithms using the grid method that divides and conquers multiple independent small problems so that the use of shuffle operations are minimized.

        \subsubsection{Enhanced Distributed Node Occlusion} \label{s:sec3.2.1}    
            \autoref{fig:1} (A) shows the overall pipeline of the enhanced node occlusion evaluation algorithm. First, it starts with a given dataframe $D_{pos}$ containing vertices $v_{i}$ and its $xy$-coordinate $p_{i}=(x_i, y_i)$. This dataframe can be viewed as vertices placed in a two-dimensional plane with their boundaries (A-\encircle{a}). Each vertex has its boundary which is represented as yellow circles in A-\encircle{a} with the same radius. In order to count cases where boundaries are overlapping each other (i.e., A-\encircle{b}) without \textit{join} operation, grid division (A-\encircle{1}) is conducted. The size of each grid is $2r$ by $2r$ where $r$ denotes the radius of each boundary. By setting grid size to $2r$ square, each vertex's potential occlusions are all located in adjacent 9 grids including its own grid. To compare each potential occlusion, each vertex is mapped to each grid where its boundary is overlapping. As a result of this process, it now has dataframe containing grid ids and classified vertices for each grid (A-\encircle{2}). Next, it applies \textit{group-by} operation on the grid id column, and exact $O(n^2)$ pair-wise comparison is performed for each group with aggregate function for exploding all vertices pairs. This gives us dataframe with vertices pairs overlapping each other including duplicated pairs. Finally, the \textit{distinct} operation is performed on the dataframe to remove duplicated pairs (A-\encircle{3}). The number of rows in the resulting dataframe is $N_c$.

        \subsubsection{Enhanced Distributed Edge Crossing} \label{s:sec3.2.2}    
            \autoref{fig:1} (B) shows the overall pipeline of the enhanced edge crossing evaluation algorithm. First of all, it starts with given dataframes $D_{pos}$ and Edge Dataframe which contains vertex id pairs of each edge. And it generates a new dataframe containing each edge's two vertex ids and their $xy$-coordinates in one row by performing the equal joining $D_{pos}$ with \textit{Edge Dataframe} on the vertex id column (B-\encircle{1}). The resulting dataframe can be seen as vertices and edges placed in a two-dimensional plane (B-\encircle{a}). In order to count cases where edges are crossing each other (i.e., B-\encircle{b}) without \textit{join} operation, grid division (B-\encircle{2}) is also conducted with some small width size $l$. But unlike the node occlusion, it divides only vertically to minimize non-comparable pairs. We define two line segments are comparable when both edges have more than one common vertical lines that they're crossing. If two line segments $s_1$ and $s_2$ are comparable at vertical lines $l_1$ and $l_2$, they are considered to be crossed if and only if their relationship between the $y$ coordinates lies on each line is reversed from $l_1$ to $l_2$. However, if we divide into grids as same as the node occlusion, we can face various situations where two line segments are non-comparable which means they don't have more than one common vertical grid lines such as a line segment crossing the top of the grid line and right of the grid line, etc. By dividing only vertically, we can minimize such cases and maximize comparable cases at the same time. In order to further minimize non-comparable cases, the grid's width size $l$ needs to be smaller. Now, edges are divided into smaller line segments for each grid. And it performs \textit{group-by} operation on each grid, and $O(n\log{n})$ edge crossing counting algorithm is conducted for each group (B-\encircle{3}). The edge counting algorithm uses two data structures to achieve $O(n\log{n})$ edge crossing counting. A sorted array $L$ consisted of the left side's $y$-coordinates $l_i$ and an initially empty balanced binary tree $R$ manages the right side's $y$-coordinates $r_i$ in non-decreasing order. Since we're only considering cases where every line segment in a group are comparable on the group's left and right grid lines, it only need to manage $y$-coordinates that each line segment is crossing with the grid lines. It sweeps through the $l_i$ of the left grid line in non-decreasing order using $L$ and updates $R$ with the new $r_i$ of the currently searching line segment $i$. We can notice that the number of line segments that cross with the currently searching line segment $i$ is the same as the number of the right side's $y$-coordinates that are greater than $r_i$ since they are reversed from the left grid to the right grid with the line segment $i$. For instance, B-\encircle{c} and B-\encircle{d} indicate line segments that are crossing with the currently searching line segment (red lines). Grey lines indicate not yet searched line segments that are not contained in $R$. Because $R$ is a balanced binary tree, it can binary search to find the number of line segments that are greater than $r_i$ and achieve $O(n\log{n})$ time complexity. As a result of this process, it now has a dataframe containing grid ids and the number of crossing lines in each grid (B-\encircle{4}). Finally, the aggregate function for summing up counted values is applied which will return the value of $E_c$.

        \subsubsection{Enhanced Distributed Edge Crossing Angle} \label{s:sec3.2.3}
            \autoref{fig:1} (C) shows the overall pipeline of the enhanced edge crossing angle evaluation algorithm. The beginning of this algorithm is the same as the enhanced edge crossing algorithm as shown in \autoref{fig:1} B-\encircle{1} and B-\encircle{2}. After dividing edges into line segments, it uses a sorted array $L$ to sweep the left grid side's $y$-coordinates $l_i$ as same as the enhanced edge crossing algorithm for each group (C-\encircle{3}). But, it uses a 2-dimensional dynamic segment tree as $R$ to manage the right grid side instead of a balanced binary tree. $R$ is updated by two factors that consisting each dimension of the $R$: angle $\theta_i$ and $y$-coordinate lies on the right grid side $r_i$. $\theta_i\in[0, \pi)$ indicates the angle between a line segment $i$ and $x$-axis. For the currently searching line segment $i$, we can group one of the crossing line segments $j$ into one of the 8 angle categories (C-\encircle{a} $\sim$ C-\encircle{h}). Each angle category has its angle range relative to the $\theta_i$ as follows:

            \begin{itemize}
              \item C-\encircle{a} left inner less ($LIL$): $[\theta_i, \theta_i + \vartheta)$
              \item C-\encircle{b} left inner greater ($LIG$): $[\theta_i + \vartheta, \theta_i + \frac{\pi}{2})$
              \item C-\encircle{c} left outer greater ($LOG$): $[\theta_i + \frac{\pi}{2}, \theta_i + \pi - \vartheta)$
              \item C-\encircle{d} left outer less ($LOL$): $[\theta_i + \pi - \vartheta, \theta_i + \pi)$
              \item C-\encircle{e} right inner less ($RIL$): $[\theta_i - \vartheta, \theta_i)$
              \item C-\encircle{f} right inner greater ($RIG$): $[\theta_i - \frac{\pi}{2}, \theta_i - \vartheta)$
              \item C-\encircle{g} right outer greater ($ROG$): $[\theta_i - \pi + \vartheta, \theta_i - \frac{\pi}{2})$
              \item C-\encircle{h} right outer less ($ROL$): $[\theta_i - \pi, \theta_i - \pi + \vartheta)$
            \end{itemize}
            
            Where $\vartheta$ denotes the ideal angle. And using the sum of each category, we can compute the edge crossing angle for $i$ as \autoref{eq:anglecaregories}.
            
            \begin{equation}\label{eq:anglecaregories}
            \begin{split}
            \sum_{e_j\in c(e_i)}{\frac{\left | \vartheta - \theta_{e_i,e_j} \right |}{\vartheta}} & = \vartheta |LIL| - (\sum{LIL} - \theta_i |LIL|)\\
            & + (\sum{LIG} - \theta_i |LIG|) - \vartheta |LIG|\\
            & + (\theta_i |LOG| - (\sum{LOG} - \pi |LOG|)) - \vartheta |LOG|\\
            & + \vartheta |LOL| - (\theta_i |LOL| - (\sum{LOL} - \pi |LOL|))\\
            & + \vartheta |RIL| - (\theta_i |RIL| - \sum{RIL})\\
            & + (\theta_i |RIG| - \sum{RIG}) - \vartheta |RIG|\\
            & + ((\sum{ROG} + \pi |ROG|) - \theta_i |ROG|) - \vartheta |ROG|\\
            & + \vartheta |ROL| - ((\sum{ROL} + \pi |ROL|) - \theta_i |ROL|)
            \end{split}
            \end{equation}
            
            If each angle group contains only angles that all of their corresponding segment $j$ are satisfying $r_j$ > $r_i$, we can compute the edge crossing angle for currently searching line segment $i$ by using \autoref{eq:anglecaregories}. Since $R$ is a 2-dimensional dynamic segment tree with angle and $y$-coordinate dimension, we can get each angle group's cardinality and summation value with $y$-coordinate condition $r_j$ > $r_i$ with $O(n\log^2{n})$ time complexity. For instance, C-\encircle{i} indicates line segments located in each angle group for the currently searching line segment (red line). Grey lines indicate not yet searched line segments. As a result of this step, it has a dataframe containing grid ids and the number of crossing line segments with the sum of crossing angles of the corresponding grid (C-\encircle{4}). Finally, the aggregate function for summing up counted values and crossing angles is applied so that it can directly compute $E_{ca}$.

\section{Experiments}

\begin{table}
    \small
    \caption{The number of vertices and edges of each dataset} \label{table:datasets}
    \begin{tabularx}{\linewidth}{r|rrl}
        \toprule
        \textbf{Dataset} & $|V|$ & $|E|$ & Description \\ \midrule 
        \textbf{ego-Facebook} & 4,039 & 88,234 & Facebook social network \\
        \textbf{musae-facebook} & 22,470 & 171,002 & Facebook page network \\ 
        \textbf{musae-github} & 37,700 & 289,003 & Github social network \\
        \textbf{soc-RedditHyperlinks} & 35,776 & 286,561 & Reddit hyperlinks network \\ 
        \textbf{cit-HepTh} & 27,770 & 352,807 & Arxiv citation network \\
        \textbf{soc-Epinions1} & 75,879 & 508,837 & Online social network \\ 
        \bottomrule
    \end{tabularx}
\end{table}

\begin{table}
    \centering
    \caption{Computational time in seconds.} \label{table:computationaltime}
    \resizebox{1.0\textwidth}{!}{
        \begin{tabularx}{1.45\linewidth}{cc|rrrrrr}
            \toprule
             & & ego-Facebook & musae-facebook & musae-github & soc-RedditHyperlinks & cit-HepTh & soc-Epinions1 \\ \midrule 
            \multirow{5}{*}{\textbf{Greadability.js}} & \textbf{$N_c$} & 0.3 & 8 & 24 & 23 & 13 & 103 \\
            & \textbf{$M_a$} & 0.4 & 0.6 & 1 & 0.5 & 1 & 1 \\ 
            & \textbf{$M_l$} & 0.02 & 0.2 & 0.07 & 0.06 & 0.09 & 0.9 \\
            & \textbf{$E_c$} & 339 & 1,828 & 7,540 & 6,107 & 13,771 & 52,545 \\
            & \textbf{$E_{ca}$} & 339 & 1,828 & 7,540 & 6,107 & 13,771 & 52,545 \\ 
            \midrule 
            \multirow{5}{*}{\textbf{Spark exact}} & \textbf{$N_c$} & 4 & 14 & 43 & 36 & 22 & 160 \\
            & \textbf{$M_a$} & 6 & 4 & 7 & 3 & 4 & 8 \\ 
            & \textbf{$M_l$} & 4 & 3 & 4 & 2 & 3 & 5 \\
            & \textbf{$E_c$} & 792 & 2,988 & 8,641 & 8,482 & 12,483 & 27,115 \\
            & \textbf{$E_{ca}$} & 882 & 3,367 & 9,129 & 8,813 & 13,443 & 30,178 \\ 
            \midrule 
            \multirow{3}{*}{\textbf{Enhanced algorithm}} & \textbf{$N_c$} & 3 & 2 & 2 & 5 & 2 & 6 \\
            & \textbf{$E_c$} & 35 & 64 & 131 & 124 & 129 & 359 \\
            & \textbf{$E_{ca}$} & 234 & 421 & 1,025 & 1,047 & 1,294 & 1,668 \\ 
            \bottomrule
        \end{tabularx}
    }
\end{table}

\begin{figure}
    \centering
    \includegraphics[width=0.6\columnwidth]{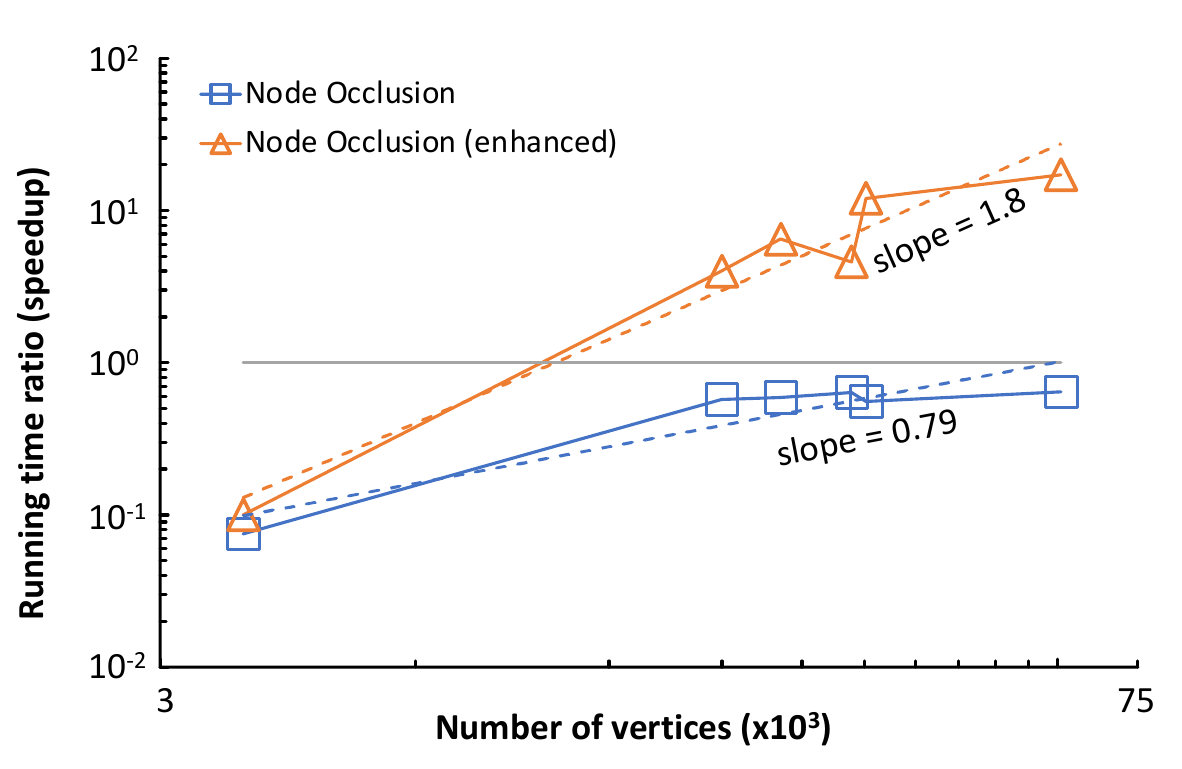}
    \caption{Running time ratio by the number of vertices. Only readability evaluation algorithms whose running time is influenced by the number of vertices are shown. The dotted lines indicate fitted power functions. The grey line indicates $1\times$ improvement where running time becomes the same as $Greadability.js$.}
    \label{fig:scalabilityplotv}
\end{figure}

\begin{figure}
    \centering
    \includegraphics[width=0.6\columnwidth]{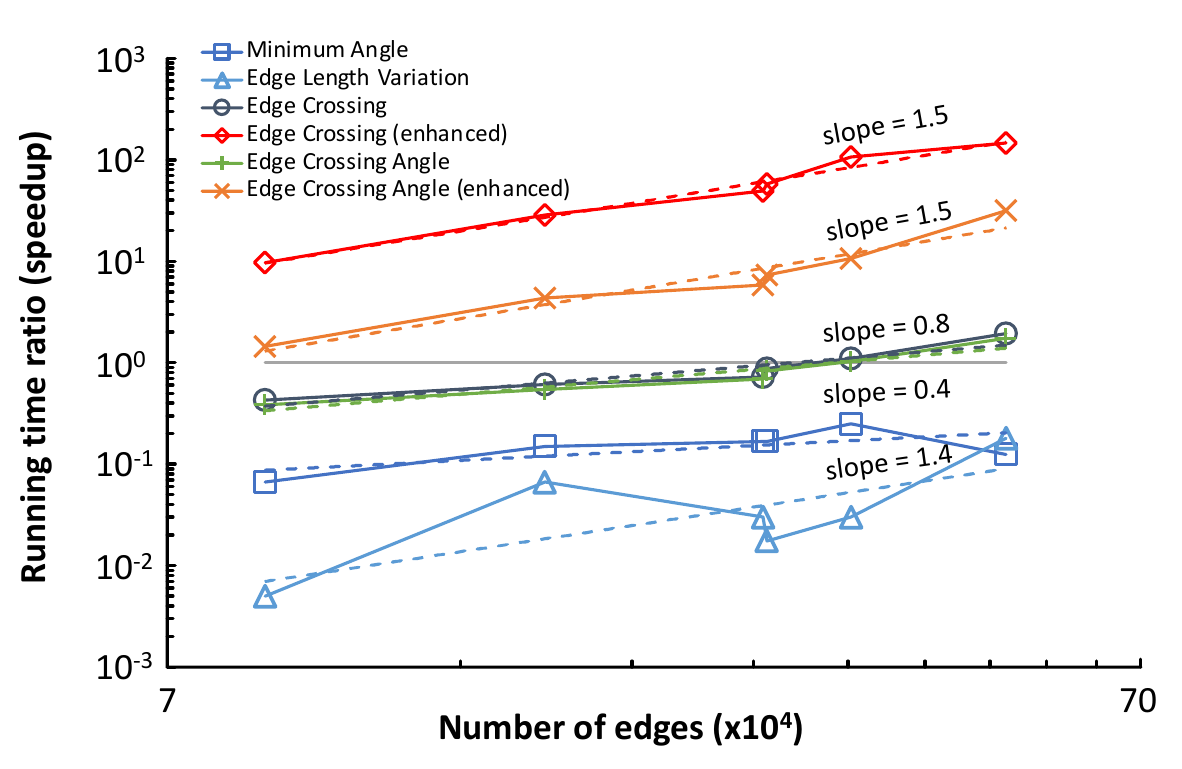}
    \caption{Running time ratio by the number of edges. Only readability evaluation algorithms whose running time is influenced by the number of edges are shown. The dotted lines indicate fitted power functions. The grey line indicates $1\times$ improvement where running time becomes the same as $Greadability.js$.}
    \label{fig:scalabilityplote}
\end{figure}
        
\begin{table}
\centering
\caption{Percentage errors of the enhanced algorithms on random layouts of each dataset. Node occlusion proved its exactness by showing 0\% error rates on all datasets. Edge crossing and Edge crossing angle show an average of about 1.5\% and 4.5\% error rates respectively.} \label{table:errors}
\resizebox{\columnwidth}{!}{
\begin{tabular}{c|cccccc}
\toprule
\textbf{Dataset} & ego-Facebook & musae-facebook & musae-github & cit-HepTh & soc-RedditHyperlinks & soc-Epinions1 \\ \midrule 
\textbf{$N_c$} & 0.0\% & 0.0\% & 0.0\% & 0.0\% & 0.0\% & 0.0\% \\
\textbf{$E_c$} & 1.4\% & 1.5\% & 1.5\% & 1.5\% & 1.4\% & 1.4\% \\
\textbf{$E_{ca}$} & 4.8\% & 1.0\% & 7.9\% & 5.2\% & 3.8\% & 4.4\%  \\ 
\bottomrule
\end{tabular}
}
\end{table}

\begin{table}
\centering
\caption{Mean values and standard deviations of the percentage errors of the edge crossing across 10 different layouts generated by using the Fruchterman-Reingold layout algorithm of the ego-Facebook dataset.} \label{table:FRlayoutserrors}
\begin{tabular}{cc|ccc}
\toprule
\textbf{grid size} & \textbf{grid orientation} & \textbf{mean} & \textbf{std} \\ \midrule 
\multirow{3}{*}{0.10} & vertical & 4.5\% & 0.032 \\
 & horizontal & 6.1\% & 0.042 \\
 & both & 4.2\% & 0.032 \\
\midrule 
\multirow{3}{*}{0.05} & vertical & 2.5\% & 0.019 \\ 
 & horizontal & 3.4\% & 0.024 \\
 & both & 2.4\% & 0.018 \\
\bottomrule
\end{tabular}
\end{table}

\begin{figure}
    \centering
    \includegraphics[width=0.8\columnwidth]{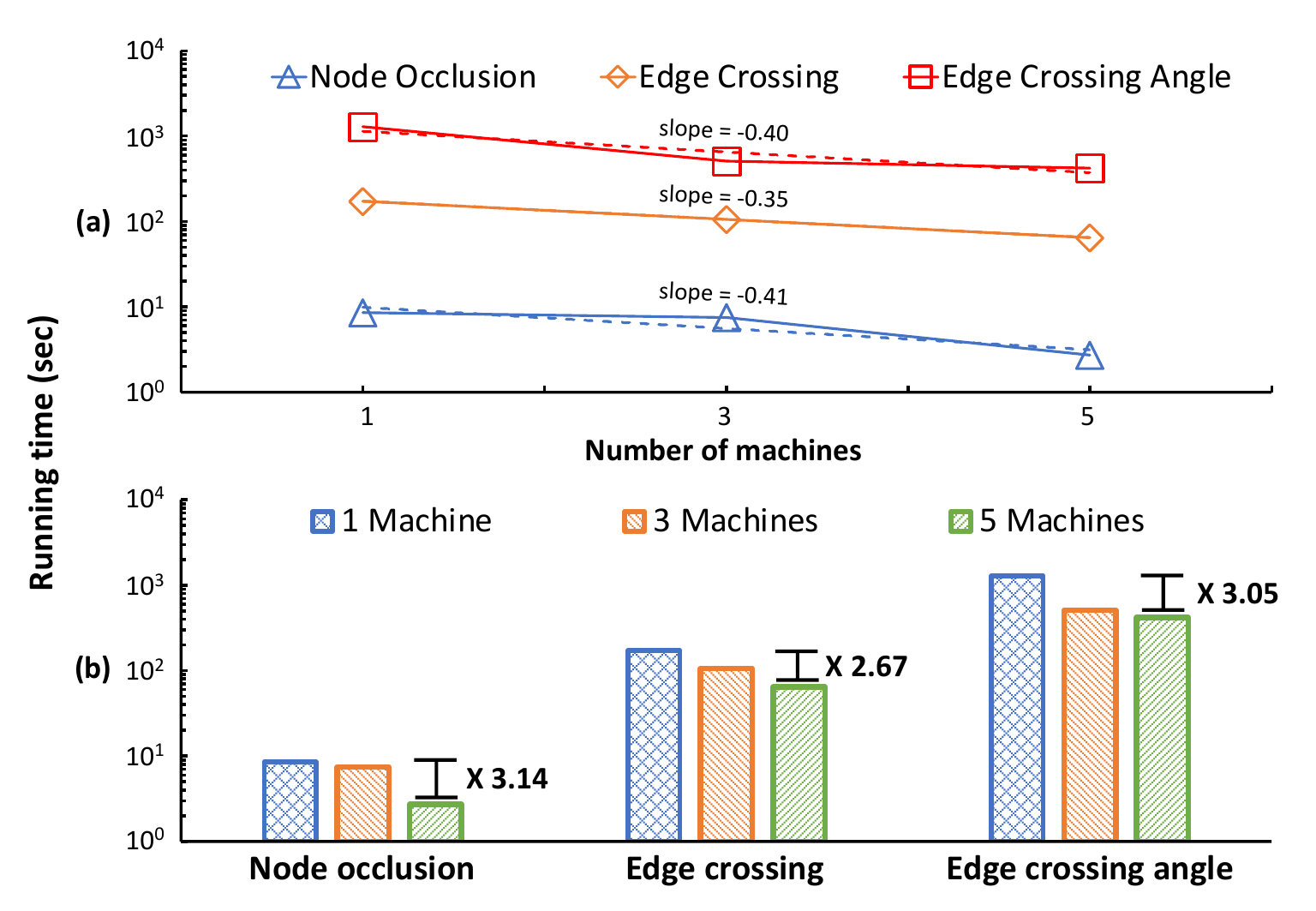}
    \caption{Strong scalability of proposed readability evaluation algorithms on the musae-facebook dataset. The dotted lines on (a) indicate fitted exponential functions.}
    \label{fig:scalabilityplot}
\end{figure}

We conducted quantitative experiments to evaluate the scalability and accuracy of our exact and enhanced algorithms.

\noindent
\textbf{Datasets.} Six datasets from SNAP~\cite{snapnets} were used, with vertex counts from 4K to 75K and edge counts from 88K to 508K (\autoref{table:datasets}).

\noindent
\textbf{Competitors.} For Minimum Angle, Edge Crossing, and Edge Crossing Angle, we compared our algorithms against Greadability.js~\cite{gove2018pays}, the only available implementation. For metrics not provided by Greadability.js (Node Occlusion and Edge Length Variation), we implemented single-machine algorithms in JavaScript.

\noindent
\textbf{Environments.} Our algorithms were tested on Google Cloud Platform Dataproc with six machines (n1-standard-8: 8 vCPUs, 32 GB RAM, 128 GB disk each), while Greadability.js ran on an Intel Core i7-7700 CPU @ 3.60GHz with 64GB RAM.

\subsection{Experiment 1: Running Time Comparison}
\noindent
\textbf{Setup:} We measured the running times of Greadability.js, exact algorithms, and enhanced algorithms on random layouts for each dataset, with vertices randomly placed within $x, y \in [0, 100]$.

\noindent
\textbf{Results:} \autoref{table:computationaltime} shows running times across algorithms. Greadability.js computes $E_c$ and $E_{ca}$ together, resulting in identical times for these metrics. \autoref{fig:scalabilityplotv} and \autoref{fig:scalabilityplote} show time ratios relative to Greadability.js by vertex and edge count, respectively. In \autoref{fig:scalabilityplotv}, enhanced Node Occlusion achieves up to $17\times$ speedup, while the exact version remains below $1\times$. In \autoref{fig:scalabilityplote}, enhanced algorithms achieve up to $146\times$ improvement in Edge Crossing and $31\times$ in Edge Crossing Angle. Exact algorithms require larger graphs for significant speedups, while enhanced algorithms show substantial improvements on smaller graphs.

\subsection{Experiment 2: Accuracy Analysis}
\noindent
\textbf{Setup:} To test accuracy, we measured readability metrics using our enhanced algorithms on random layouts and layouts generated with the Fruchterman-Reingold algorithm. Ground-truth values for each metric were computed using straightforward C++ implementations.

\noindent
\textbf{Results:} \autoref{table:errors} shows the percentage errors for each dataset. Node Occlusion yielded 0\% error as expected. Edge Crossing and Edge Crossing Angle showed averages of 1.5\% and 4.5\% error, respectively—significantly lower than the deep learning approach \cite{haleem2019evaluating}, which reported errors of up to 22.20\% and 55\%. Accuracy for Edge Crossing and Angle decreases with shorter edge lengths, as these increase non-comparable pairs. We tested Edge Crossing on 10 Fruchterman-Reingold layouts of the ego-Facebook dataset under different grid configurations (see \autoref{table:FRlayoutserrors}). Reducing grid size and selecting maximum values across both grid orientations improved accuracy. Despite slight increases in error for layout-generated graphs, accuracy remains much higher than prior methods.

\subsection{Experiment 3: Scalability Analysis}
\noindent
\textbf{Setup:} To assess scalability, we measured running times of our enhanced algorithms on the musae-facebook dataset with varying machine counts.

\noindent
\textbf{Results:} \autoref{fig:scalabilityplot} shows strong scalability, with enhanced Node Occlusion and Edge Crossing Angle achieving a slope of about -0.4, meaning doubling machines reduces running time by $2^{0.4} \approx 1.31\times$. All enhanced algorithms showed up to $3.14\times$ speedup as machine counts increased, demonstrating effective scalability for large datasets.
\section{Conclusion}

The lack of scalable and accurate evaluation algorithms limits our ability to effectively analyze large graph layouts. To address this, we introduced two scalable readability evaluation algorithms—exact and enhanced versions—designed for distributed environments. Our experiments demonstrate that these algorithms offer substantial improvements in running time, accuracy, and scalability for large-scale graphs compared to single-machine approaches. Additionally, we highlighted the practical applicability of our methods through an application in layout optimization, underscoring their value for handling complex graph analysis tasks efficiently.

\bibliographystyle{unsrtnat}
\bibliography{reference}

\end{document}